\documentclass[aps,prl,twocolumn,letterpaper,showpacs,preprintnumbers,amsmath,amssymb]{revtex4}

\usepackage{graphicx}
\usepackage{dcolumn}
\usepackage{bm}

\begin{document}

\title{Impairment of double exchange mechanism in electron transport of iron pnictides}

\author{Lei Hao, Chi-Cheng Lee, and T. K. Lee}
 \address{Institute of Physics, Academia Sinica, Nankang, Taipei 11529, Taiwan}

\date{\today}

\begin{abstract}
Double exchange mechanism is believed to favor transport along
ferromagnetic directions, the failure of which in explaining
the unusual resistivity anisotropy in iron pnictides is
investigated. Several factors intrinsic to the microscopic
mechanism of transport in iron pnictides are identified and
analyzed, including the moderate Hund's coupling, low local
moment, and presence of two anisotropic degenerate orbitals
$xz$ and $yz$. In particular, the substantial second neighbor
hoppings are found to be decisive in giving results opposite to
the double exchange picture. In high temperature nonmagnetic
phase, orbital ordering is shown to give the right trend of
resistivity anisotropy as observed experimentally, advocating
its essential role in electron transport of iron pnictides.
\end{abstract}


\maketitle


Close interplay among spin, orbital, and lattice degrees of
freedom together with the high-temperature superconductivity
make iron pnictides a rich playground to study the effects of
electron correlations. Unlike cuprates and manganites, all
families of iron pnictides are metallic without doping. This
greatly interests researchers in studying the transport
property under different circumstances and studying its
relationship to all kinds of degrees of freedom
\cite{chu10,yi11,dusza10,ying10,kuo11,tanatar10,blomberg11,fisher11}.
Recently, an unusual resistivity anisotropy was observed in
detwined underdoped Ba(Fe$_{1-x}$Co$_{x}$)$_{2}$As$_{2}$
\cite{chu10,yi11,dusza10,ying10,kuo11,tanatar10} and was
confirmed in other iron pnictides \cite{blomberg11,fisher11}.
The low temperature normal state of the material is an
antiferromagnetic (AF) metal with collinear ($\pi,0$) ordering.
Experimentally, the resistivity was found to be larger along
the ferromagnetic (F) direction than along the AF direction.
This is thought to be in direct contradiction to at least two
reasonings: (1) The ferromagnetic Hund's coupling between local
moments and the itinerant electrons indicates a double exchange
(DE) like correlation between spin configuration and transport,
which implies a lower resistivity along the F direction
\cite{chu10,turner09}. (2) The scattering rate is expected to
be larger along the AF direction \cite{valenzuela10}, which is
confirmed in optical conductivity experiment
\cite{dusza10,sanna11}. Another surprising result is that the
resistivity anisotropy is largest close to the border between
the AF phase and the superconducting dome, where the magnetism
and structural distortion are weaker \cite{chu10}.

Though conceptually puzzling, the early experimental results
mentioned above are reproduced by several theoretical
calculations, including both mean field studies
\cite{valenzuela10,sugimoto10,lv11} and other more advanced
techniques such as the dynamical mean field theory
\cite{yin10arx,laad10,fernandes11}. Nevertheless, a clear
physical picture concerning why the DE mechanism not working
has not been addressed. Current understanding of mean field
results on resistivity anisotropy is only based on the topology
and morphology of the Fermi surface \cite{valenzuela10},
leaving the underlying mechanism untold. The failure of the
interplay between ferromagnetism and conductivity in the AF
metallic phase is thus hazy and still remains a puzzle. If the
scattering rate is indeed stronger along the AF direction
helping the electron transport along F direction
\cite{dusza10}, there must be a non-negligible competitor
making the DE mechanism much less effective. Since several
calculations via different methods all observe this same
unusual resistivity anisotropy, the underlying physics should
be simple. It is thus interesting and timely to explore the
applicability of the DE mechanism in iron pnictides, especially
because the DE based model itself has been used successfully to
unify the magnetic correlations in iron-based superconductors
\cite{yin10} and account for the novel spin dynamics
\cite{lv10}. Besides, the reason of impairment of double
exchange mechanism in transport may also happen in other
materials. In addition, orbital ordering is known to play an
important role in many aspects of the system
\cite{lee09,lv09,kontani10,chen10}. Surprisingly, mean field
calculations have shown that orbital ordering is irrelevant to
\cite{sugimoto10} or even give opposite trend compared to
\cite{valenzuela10} the experimental results. It is extremely
desirable to reexamine the correlation between orbital ordering
and the resistivity anisotropy presently.

In this Letter, we report a theoretical study on the
applicability of DE mechanism in the electron transport of iron
pnictides. A so-called orbital-degenerate DE model
\cite{yin10,lv10} is used to uncover the underlying mechanism
of resistivity anisotropy both below and slightly above the
structural and magnetic phase-transition temperatures. Although
the DE effect is already weaken by the small Hund's coupling,
low local moments and two degenerate anisotropic orbitals, the
relation between magnetism and transport anisotropy is still
expected to follow the DE idea. The \emph{substantial
second-neighbor hoppings} are found to give rise to both a
gapping feature in the band structure and a frustration in
kinetic energy gain in transport along the F direction in the
AF phase and is identified as an important factor in
competition with the DE mechanism. In the high-temperature
nonmagnetic phase, the orbital ordering that could arise from a
tiny external stress is shown to give the right trend of
resistivity anisotropy and thus contributes positively to the
experimental resistivity anisotropy.


The model $H$=$H_{0}$+$H_{K}$+$H_{S}$+$H_{L}$ contains four
terms. $H_{0}$ is the tight binding term up to second neighbor
hopping \cite{raghu08,yin10}. $H_{K}$ is the local Hund's
coupling between localized spin and the itinerant electron
spin, and is defined as
$H_{K}$=$-\frac{1}{2}K\sum_{\mathbf{i},\mu,\alpha,\beta}
\mathbf{S}_{\mathbf{i}}\cdot(d^{\dagger}_{\mathbf{i},\mu,\alpha}
\boldsymbol{\sigma}_{\alpha\beta}d_{\mathbf{i},\mu,\beta})$.
$\mu$ indicates the orbital index. $K$$>$$0$ is the strength of
Hund's coupling, $\sigma_{i}$ is the $i$-th Pauli's matrix
acting in the spin subspace, and $\mathbf{S_{\mathbf{i}}}$ is
the localized spin on site $\mathbf{i}$. In order to maximize
the DE mechanism, we take $\mathbf{S_{\mathbf{i}}}$ as
classical spins \cite{yin10}. Since we focus on the
$\mathbf{Q}$=$(\pi,0)$ AF phase, we can take
$\mathbf{S}_{\mathbf{i}}$=$S\hat{z}\cos(\mathbf{Q}\cdot\mathbf{R}_{\mathbf{i}})$.
In the following, we redefine $J$=$KS/2$ to measure the
strength of Hund's coupling. The first two terms of the model
constitute the DE model in the limit of large Hund's coupling,
which is well-known to facilitate transport along F
directions\cite{anderson55,dagotto01}.

$H_{S}$ is the superexchange term between localized spins,
which competes with the first two terms to determine the ground
state magnetic order and spin dynamics \cite{yin10,lv10}. Since
we concentrate on the electron transport in the fixed
$\mathbf{Q}=(\pi,0)$ AF phase in this work, $H_{S}$ is
neglected in what follows. The final term is a crystal field
splitting between the two orbitals as a result of the external
stress field. It is written as
$H_{L}=\frac{1}{2}\Delta\sum_{\mathbf{i}}(n_{\mathbf{i},1}-n_{\mathbf{i},2})$,
where
$n_{\mathbf{i},\mu}$=$\sum_{\sigma}d^{\dagger}_{\mathbf{i},\mu,\sigma}d_{\mathbf{i},\mu,\sigma}$
is the number operator for electrons in orbital $\mu$ and on
site $\mathbf{i}$. The lattice distortion also changes the
hopping amplitudes along $x$ and $y$ directions \cite{hao07}.
However, we consider this as a minor effect in our present
system of undoped or underdoped iron pnictides and neglect it
in the following analyses.

In the $(\pi,0)$ AF phase, every unit cell contains two irons.
Relabeling the electron annihilation operators on the two
sublattices as $a_{\mathbf{k},\mu,\sigma}$ and
$b_{\mathbf{k},\mu,\sigma}$. Take the $\sigma$ spin basis as
$\psi^{\dagger}_{\mathbf{k}\sigma}$=$\{a^{\dagger}_{\mathbf{k},1,\sigma},
b^{\dagger}_{\mathbf{k},1,\sigma},a^{\dagger}_{\mathbf{k},2,\sigma},
b^{\dagger}_{\mathbf{k},2,\sigma}\}$. Ignoring $H_{S}$, the
original model is written as
$H$=$H_{0}+H_{K}+H_{L}$=$\sum_{\mathbf{k}\sigma}\psi^{\dagger}_{\mathbf{k}\sigma}
h^{\sigma}_{\mathbf{k}}\psi_{\mathbf{k}\sigma}$, where
\begin{equation}
h^{\sigma}_{\mathbf{k}}=
\begin{pmatrix} f^{\sigma}_{1a}(\mathbf{k}) & f_{3}(\mathbf{k}) &
0 & f_{5}(\mathbf{k}) \\ f_{3}(\mathbf{k}) & f^{\sigma}_{1b}(\mathbf{k}) & f_{5}(\mathbf{k}) & 0 \\
0 & f_{5}(\mathbf{k}) & f^{\sigma}_{2a}(\mathbf{k}) & f_{4}(\mathbf{k}) \\ f_{5}(\mathbf{k}) & 0 &
f_{4}(\mathbf{k}) & f^{\sigma}_{2b}(\mathbf{k}) \end{pmatrix},
\end{equation}
in which
$f^{\sigma}_{1a}(\mathbf{k})=-2t_{1}\cos(k_{y}c_{2})+\frac{\Delta}{2}-\sigma
J$,
$f^{\sigma}_{1b}(\mathbf{k})=-2t_{1}\cos(k_{y}c_{2})+\frac{\Delta}{2}+\sigma
J$,
$f^{\sigma}_{2a}(\mathbf{k})=-2t_{2}\cos(k_{y}c_{2})-\frac{\Delta}{2}-\sigma
J$,
$f^{\sigma}_{2b}(\mathbf{k})=-2t_{2}\cos(k_{y}c_{2})-\frac{\Delta}{2}+\sigma
J$,
$f_{3}(\mathbf{k})=-2t_{2}\cos(k_{x}c_{1})-4t_{3}\cos(k_{x}c_{1})\cos(k_{y}c_{2})$,
$f_{4}(\mathbf{k})=-2t_{1}\cos(k_{x}c_{1})-4t_{3}\cos(k_{x}c_{1})\cos(k_{y}c_{2})$,
$f_{5}(\mathbf{k})=-4t_{4}\sin(k_{x}c_{1})\sin(k_{y}c_{2})$.
$c_{1}$ ($c_{2}$) is lattice constant along $x$ ($y$). Along
($k_{x}$, 0) or (0, $k_{y}$), $f_{5}(\mathbf{k})$=0, the two
eigen-energies for the two orbitals are written compactly as
\begin{equation}
\lambda_{\mu,\pm}=-2t_{\mu}\cos(k_{y}c_{2})-(-1)^{\mu}\frac{\Delta}{2}\pm\sqrt{J^{2}+f^{2}_{\mu+2}(\mathbf{k})},
\end{equation}
with $\mu=1$ ($\mu=2$) for the $xz$ ($yz$) orbital.
From the above expression, the two orbitals open gaps both
along the $k_{x}$ and the $k_{y}$ directions. Along $k_{x}$,
the positions of gap opening of the two orbitals are far away
from each other by a distance of $|2(t_1-t_2)-\Delta|$. For
experimentally realistic $J$, no full gap opens along $k_{x}$,
which is consistent with a topological argument by Ran \emph{et
al} \cite{ran09}. This Dirac cone structure appears along
$k_{x}$ as an accidental crossing of the energy bands
corresponding separately to the two orbitals. However, it could
be proved that for $\Delta$=0 and small $J$ the gap positions
for the two orbitals always coincide with each other along
$k_{y}$ and are centered around $t_{1}t_{2}/t_{3}$. This
orbital-cooperated gap would highly enhance the corresponding
resistivity.

For the hopping integrals, we consider mainly the parameters
proposed by Raghu \emph{et al} \cite{raghu08} and fix the
largest hopping integral as $t_{1}$=0.4 eV \cite{lee09,yin10}.
The ratios among the four hopping integrals
$t_{1}$:$t_{2}$:$t_{3}$:$t_{4}$ are 1.3:(-1):(-0.85):(-0.85).
Results for another set of parameters used by Yin \emph{et al}
\cite{yin10}, in which
$t_{1}$:$t_{2}$:$t_{3}$:$t_{4}$=4:1.3:(-2.5):0.7, are found to
give qualitatively the same results. These two sets of
parameters agree qualitatively in the sense that the nearest
neighbor intra-orbital hopping integral of the $xz$ ($yz$)
orbital along the $y$ ($x$) direction $t_{1}$ is larger than
along the $x$ ($y$) direction $t_{2}$, a peculiar feature
resulting from the strong hybridization of the iron orbitals
with the $p$ orbitals of surrounding As ions \cite{lee09}.

The DE mechanism, first introduced for manganites, applies when
the ferromagnetic Hund's coupling $J$ is much larger than the
hopping amplitudes and for a partly filled band
\cite{anderson55,dagotto01}. Though the Hund's coupling for
iron pnictides is not as large, it is not obvious why a result
opposite to expectation based on DE is observed. To have a
feeling on the global behavior of the model, we show in Fig.
1(a) the resistivity ratio $\rho_{y}/\rho_{x}$ in a wide range
of $J$, at 50 K in the low temperature AF phase. Two electrons
per iron is considered for the undoped system
\cite{raghu08,kubo09,gao11}. The finite temperature resistivity
(conductivity) is obtained from standard Kubo's formula
\cite{hao10}. For low $J$ the resistivity anisotropy agrees
with experiments. But when $J$ increases to an unphysical value
for iron pnictides, the resistivity anisotropy changes to the
opposite behavior as compared to experiment, which is however
in agreement with expectation from the DE mechanism. One
peculiar feature on Fig. 1(a) is for very large $J$ the
resistivity anisotropy of the undoped system changes back to
the behavior same as in experiment, a result specific to the
undoped and extremely low doped systems and easily understood
from the Pauli's exclusion principle for electrons in a half
filled band.

Before answering the question why undoped and slightly electron
doped iron pnictides behave differently from DE predictions, we
first calculate the resistivity anisotropy for a single orbital
model with isotropic in-plane hoppings up to nearest neighbor,
that is to set $t_{1}=t_{2}$ and $t_{3}=t_{4}=0$. The results
show that even for extremely small $J$, the resistivity is
smaller along the F direction, following exactly the DE
prediction \cite{anderson55,dagotto01}. Thus, the small
magnitude of $J$ is not the crucial factor to bring about the
observed anisotropy.

\begin{figure}
\centering
\includegraphics[width=8.5cm,height=7cm,angle=0]{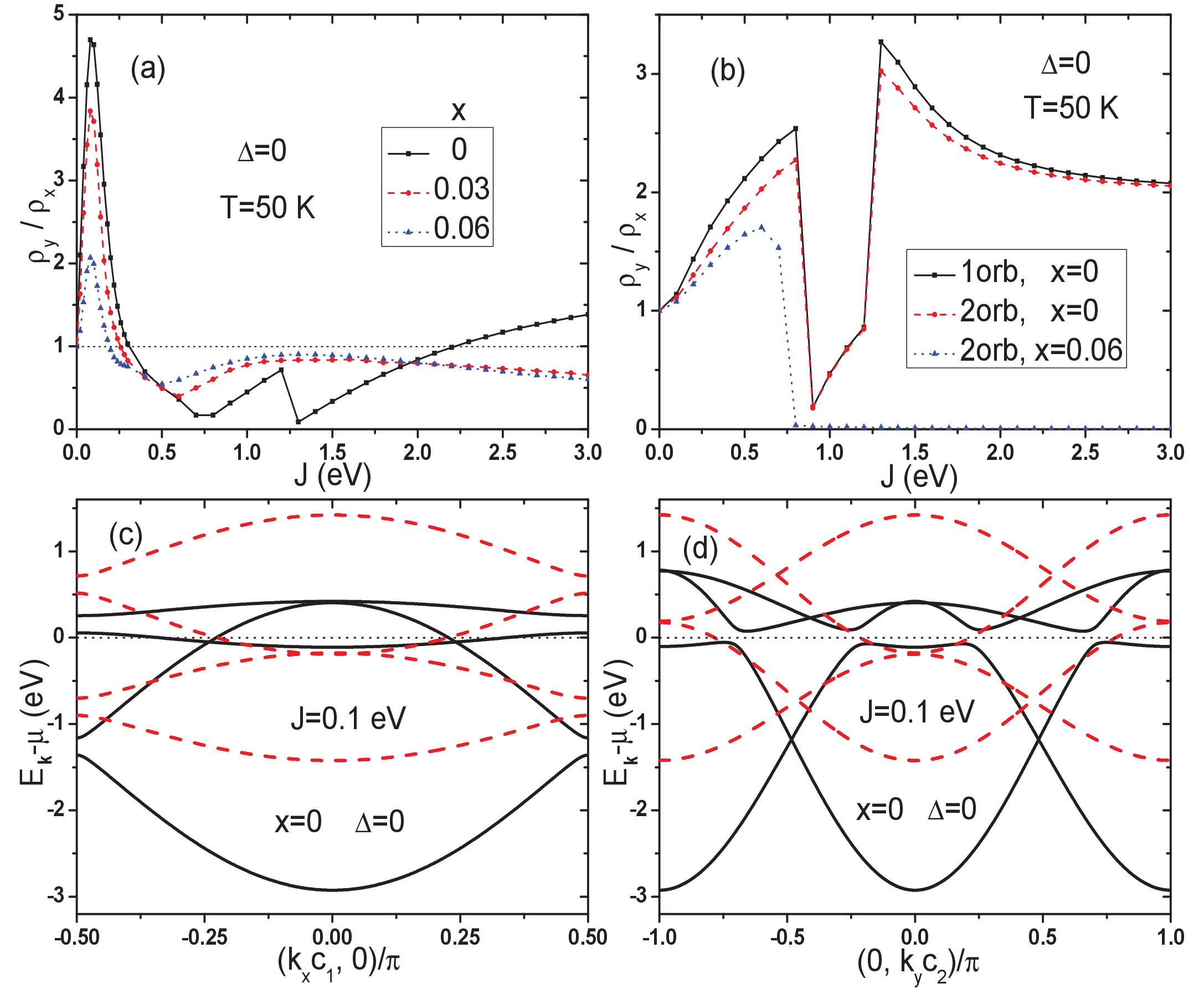}
\caption{(a) Ratio of resistivity as a function of $J$ between $\rho_{y}$ along the F direction
and $\rho_{x}$ along the AF direction, for three typical dopings at 50K.
(b) $\rho_{y}/\rho_{x}$ for isotropic one orbital (labeled as `1orb')
and two orbital (labeled as `2orb') models
with 2NN hopping, at 50 K. (c) and (d) are dispersions
along $k_{x}$ and $k_{y}$ for $J=0.1$ eV both in the presence (solid lines)
and absence (dashed lines) of 2NN hoppings, respectively.
The horizontal dotted lines in (c) and (d) mark the chemical potential.}
\end{figure}

However, the inclusion of substantial second nearest neighbor
(2NN) hoppings in combination with the small magnitude of $J$
literally induces the unusual resistivity anisotropy in iron
pnictides, which is further enhanced by the large anisotropy in
the intraorbital nearest neighbor (NN) hoppings. To see the
effect of 2NN hoppings, we show in Fig. 1(b) the resistivity
anisotropy for a model with one isotropic orbital
($t_{2}$=$t_{1}$=0.4 eV, $t_{3}$=-0.25 eV, $t_{4}$=0) and two
isotropic orbitals ($t_{2}$=$t_{1}$=0.4 eV, $t_{3}$=-0.25 eV,
$t_{4}$=0.07 eV). Though with very different line shapes, the
same qualitative behavior as in Fig. 1(a) is obtained. From
Fig. 1(b), it is also obvious that a nonzero 2NN inter-orbital
hopping $t_{4}$ tends to reduce the magnitude of the
resistivity anisotropy.

To understand the role played by 2NN hoppings, first notice
that in a square lattice the effective hopping along each of
the two in-plane directions have contributions from both the NN
and the 2NN hoppings. In the $(\pi,0)$ AF phase, the NN and 2NN
hoppings along $x$ both connect AF bonds while they connect F
and AF bonds respectively along $y$. When $J$ is very large,
the 2NN hoppings together with the NN hoppings along the AF
direction are effectively blocked, the DE mechanism simply
prevails. However, in the region of small $J$, a more intricate
competition between the maximization of kinetic energy gain and
minimization of Hund's energy appears. Since hopping along the
F bonds does not cost or gain magnetic energy, maximization of
kinetic energy gain for $J\ll t_{1}$ prefers a nearly
non-polarized on-site spin distribution. Whereas along the AF
direction, the up and down spin electrons each experiences
effectively a periodic shallow potential wells of depth $2J$,
both of two-site period and shifted from each other by one
site. This pair of periodic potential wells tends to drive the
formation of notable local spin polarizations, which can lower
the Hund's energy at the cost of kinetic energy loss. Since for
small $J$ the kinetic energy gain is dominant, for negligible
2NN hoppings the spin polarization follows largely that of the
F direction. However, the additional magnetic energy gain from
hoppings along AF 2NN bonds enhances the spin polarization, as
verified through explicit calculations. This in turn frustrates
hopping along the F bonds and finally impairs the DE for large
enough $t_{3}$. We believe that the above is the physical
picture why substantial 2NN hoppings \emph{invalidate} the DE
mechanism for very small $J$.

The effect of the 2NN hoppings could be further understood from
the band structure. For clarity, the band structures for
$J=0.1$ eV in the presence and absence of the 2NN hoppings are
shown together by solid and dashed lines in Figs. 1(c) (1(d))
along $k_{x}$ ($k_{y}$). In the absence of 2NN hoppings, the
energy bands are more metallic along $k_{y}$ than along $k_{x}$
since no energy gap opens in the whole energy range in the
former. When the 2NN hoppings are turned on, the band
structures for undoped and slightly doped systems would be
gapless along $k_{x}$ but would open a gap along $k_{y}$ as we
have mentioned following Eq.(2). This peculiar change in the
direction of gap opening tendency follows exactly the change
from DE behavior to anti-DE behavior and thus provides a very
clear picture how 2NN hoppings invalidate DE. The gap along
$k_{y}$ exists once $|t_{3}/t_{1}|\ge1/2$, providing a good
reference critical value for the 2NN hoppings.

Another important feature of iron pnictides different from
standard DE systems (e.g., manganites) is that the two relevant
orbitals are anisotropic with respect to $x$ and $y$, which is
manifested through $|t_{1}/t_{2}|>1$. Comparing Fig. 1(a) and
Fig. 1(b), this intrinsic anisotropy enhances the resistivity
anisotropy and gives rise to the peak structure in the small
$J$ region on Fig. 1(a). The low $J$ peak is far less obvious
in the absence of this anisotropy, as could be seen from Fig.
1(b).

Former mean field calculations found smaller resistivity
anisotropy for high moment states than for low moment states
\cite{valenzuela10,sugimoto10}. According to Fig. 1(a), the
reduction of resistivity anisotropy in the high moment phase
indicates that iron pnictides should belong to the region
changing from anti-DE behavior to DE behavior. We notice that
there are iron based superconductors which are considered as
having larger effective $J$, such as FeTe$_{1-x}$Se$_{x}$
\cite{yin10}. One prediction worthy of future experimental
inspection is that opposite resistivity anisotropy with larger
resistivity along the AF direction in the underdoped region of
these materials might be observed.

One common feature in iron pnictides is that a structural
transition always occurs at a higher or the same temperature as
compared with the AF transition \cite{chu10}, in which the
orbital ordering and fluctuations are considered by many to be
essential \cite{lee09,lv09,kontani10}. It is thus very
surprising that former mean-field studies showed that orbital
ordering is anti-correlated with the resistivity anisotropy in
the AF phase \cite{valenzuela10,sugimoto10}. Experimentally,
the same resistivity anisotropy is observed in the high
temperature nonmagnetic phase close to the AF and structural
transition when an external stress field is applied to induce a
lattice distortion following the structural anisotropy in the
AF phase \cite{chu10}. Here we would like to single out the
role played by the orbital ordering in this phase. In the low
doping region, we expect crystal field effect to be dominant,
and consider a dominance of the $xz$ orbital for a lattice with
a larger lattice constant along $x$, implying $\Delta<0$ in our
model \cite{hao07}. As shown in Fig. 2(a), orbital ordering
gives the \emph{right} trend of resistivity anisotropy compared
to experiment. We then calculate the orbital polarization for
states close to the chemical potential, which are most relevant
to electron transport, for a typical doping and a series of
increasing energy windows $\delta E$ in Fig. 2(b). Denoting the
integrated density of states of $xz$ ($yz$) orbital within
$\delta E$ as $\delta n_{1}$ ($\delta n_{2}$), the orbital
polarization $\delta_{12}$ is defined as $(\delta n_{1}-\delta
n_{2})/(\delta n_{1}+\delta n_{2})$. It is clear that
$\delta_{12}$ close to the chemical potential has the same
character as the total orbital polarization set by the sign of
$\Delta$. Note that fluctuations via nematic correlations
\cite{chu10,laad10} neglected here might be important to
provide a complete picture for the high temperature resistivity
anisotropy and deserves future investigations.

\begin{figure}[tbp]
\centering
\includegraphics[width=8.5cm,height=4cm,angle=0]{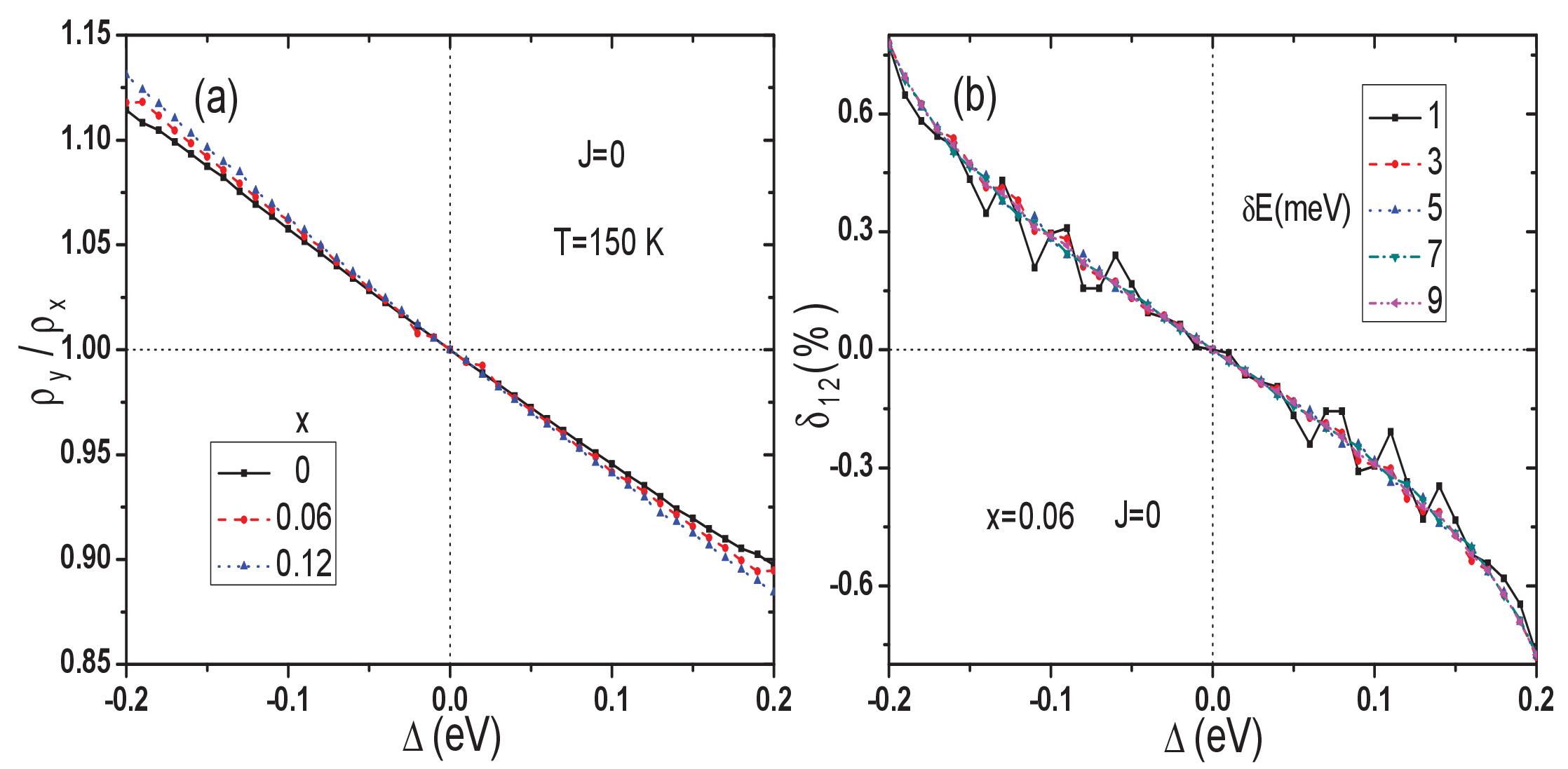}
\caption{(a) Ratio of resistivity between the direction with short lattice constant ($\rho_{y}$)
and the direction with long lattice constant ($\rho_{x}$), for three typical dopings.
Temperature is taken as 150K, without of magnetization.
(b) The orbital polarization $\delta_{12}$ (in percentage)
close to the chemical potential for a series of increasing energy windows $\delta E$.}
\end{figure}

The above correlation between orbital polarization and
resistivity anisotropy
is encoded in the intrinsic anisotropy of the two degenerate
orbitals.
As could be seen from Eq.(1) for $J=0$, the effective
intraorbital hopping along $k_{x}$ is
$-2(t_{2}+2t_{3})\cos(k_{x}c)$ for $xz$ orbital and is
$-2(t_{1}+2t_{3})\cos(k_{x}c)$ for $yz$ orbital \cite{raghu08}.
For both of two parameter sets considered,
$|t_{1}+2t_{3}|<|t_{2}+2t_{3}|$ \cite{raghu08,yin10}. So in a
qualitative sense, a dominant $xz$ polarization close to the
chemical potential implies an easier hopping and conduction
along $x$ direction. In comparison, $yz$ has a larger effective
hopping along the $y$ direction and thus facilitates the
conduction along $y$ direction.

In conclusion, the puzzling interplay between magnetic order
and electron transport in iron pnictides is identified quite
different from DE-dominated systems (e.g., manganites). The
small Hund's coupling, low local moments and intrinsic
anisotropy in the NN hoppings all make the DE a bad
description. In particular, the substantial 2NN hoppings are
found to be decisive in competing with the DE mechanism,
greatly frustrating the kinetic energy gain in transport along
the ferromagnetic direction. We have also found in the high
temperature nonmagnetic phase that, an $xz$ dominated
polarization which might arise as a result of an external
stress could give the same trend of resistivity anisotropy as
observed in experiments, advocating the possible essential role
of orbital ordering in the resistivity anisotropy.

We reemphasize that $|t_{3}/t_{1}|\sim0.5$ is a good reference
critical value of the 2NN intra-orbital hopping to impair the
DE mechanism (for small $J$), the exact value of which depends
on $t_{4}$ and $J$. First-principles 2NN hoppings for all three
$t_{2g}$ (two $e_{g}$) orbitals are larger (smaller) than half
of the corresponding NN hoppings \cite{lee09}. Since the Fermi
surface of iron pnictides are dominated by $t_{2g}$ orbitals, a
model containing more orbitals should not change our previous
arguments. Similar impairment of the DE mechanism might be
observed in other $t_{2g}$-dominated systems \cite{bhobe10}.
SrCrO$_3$ is also an AF metal with ($\pi$, $\pi$, 0) magnetic
order similar to iron pnictides \cite{qian11}. However, our
first-principles calculations give 2NN hoppings smaller than
half of NN hoppings for all five orbitals, predicting that DE
should persist in that system.  This work was supported by the
NSC Grant No. 98-2112-M-001-017-MY3. Part of the calculations
was performed in the National Center for High-Performance
Computing in Taiwan.\index{}


\end{document}